\documentclass{sigchi-ext}
\usepackage[T1]{fontenc}
\usepackage{textcomp}
\usepackage[scaled=.92]{helvet} 
\usepackage{graphicx} 
\usepackage{balance}  
\usepackage{booktabs} 
\usepackage{ccicons}  
\usepackage{ragged2e} 

\copyrightinfo{
Accepted to the Workshop on "Designing Mobile Interactions for the Ageing Populations" at CHI 2017. \\
Permission to make digital or hard copies of all or part of this work for personal or classroom use is granted without fee provided that copies are not made or distributed for profit or commercial advantage and that copies bear this notice and the full citation on the first page. \\
{\emph{CHI'17}}, May 6, 2017, Denver, USA}

\title{Designing for older adults: review of touchscreen design guidelines}

\numberofauthors{5}
\author{%
  \alignauthor{%
    \textbf{Leysan Nurgalieva}\\
    \affaddr{University of Trento} \\
    \affaddr{Trento, Italy} \\
    \affaddr{leysan.nurgalieva@unitn.it} }
    \alignauthor{%
    \textbf{Fabio Casati}\\
    \affaddr{University of Trento}\\
    \affaddr{Trento, Italy} \\
    \email{fabio.casati@unitn.it} } \vfil 
    \alignauthor{%
    \textbf{Juan Jos\'e Jara Laconich}\\
    \affaddr{University of Trento}\\
    \affaddr{Trento, Italy}\\
    \email{juan.jaralaconich@unitn.it} }
    \alignauthor{%
    \textbf{Maurizio Marchese}\\
    \affaddr{University of Trento}\\
    \affaddr{Trento, Italy}\\
    \email{maurizio.marchese@unitn.it} } \vfil 
    \alignauthor{%
    \textbf{Marcos B\'aez}\\  
    \affaddr{University of Trento}\\
    \affaddr{Trento, Italy}\\
    \email{baez@disi.unitn.it} \\ } }

\begin{document}
\maketitle
\RaggedRight{} 


\begin{abstract}
The distinct abilities of older adults to interact with computers has motivated a wide range of contributions in the the form of design guidelines for making technologies usable and accessible for the elderly population. However, despite the growing effort by the research community, the adoption of guidelines by developers and designers has been scant or not properly translated into more accessible interaction systems. 
\\In this paper we explore this issue by reporting on a qualitative outcomes of a systematic review of 204 research-derived design guidelines for touchscreen applications. We report first on the different definitions of ``elderly" and assess the reliability, organization and accessibility of the guidelines. Then we present our early attempt at facilitating the reporting and access of such guidelines to researchers and practitioners. 
\end{abstract}


\section{Introduction}
In current conditions of increasing global ageing, it is particularly important to address deteriorating abilities of older adults while designing interactive touchscreen technologies for this particular category of users. One way to achieve this is to use guidelines tailored to older users during the design process. However, due to a large body of literature on this topic, it is not always easy to clearly recognize, extract, and apply them. Moreover, quality of design findings and recommendations should be considered, as well as their consistency and validity. Therefore, there is a constant need to review, understand, and structure design guidelines making it easier to apply them to the actual design process.

\begin{marginfigure}[0pc]
  \begin{minipage}{\marginparwidth}
    \centering
  \includegraphics[width=.8\columnwidth]{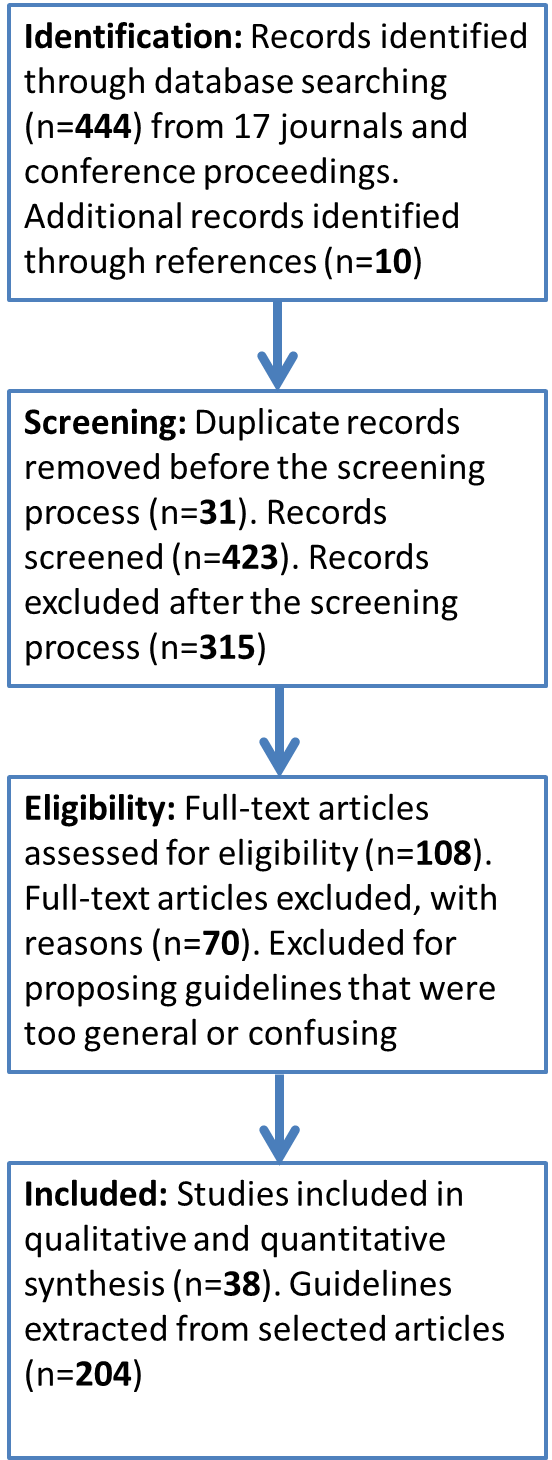}
  \captionof{figure}{PRISMA flow diagram. Process of obtaining relevant papers.}
  \label{figure:collage}
  \end{minipage}  
\end{marginfigure}

Previous research has emphasized the importance of design guidelines as precise and reliable recommendations to refer to while developing technologies for the ageing population. Early attempts to raise awareness and compile guidelines comprise the work of Smith and Mosier \cite{smith1986guidelines} with the primary focus to support general design of the user interface, however, without considering a specific technology type or older adults as a target population.

Other surveys provided general synthesis of the literature on the topic \cite{boll2015user}, as well as compilations of guidelines focusing either on specific issues related to older adults interacting with technologies, for instance, use of Social Network Services (SNSs) \cite{coelho2016literature}, or structuring them based on usability issues older adults face \cite{claypoole2016keeping}. Some of the reviews, like \cite{mi2014heuristic}, were aiming to reduce ``the gap between a designer's conceptual model and a user's mental model of the design''. However, most of the considered works were not systematic reviews, and did not specifically consider clarity, readability, and validation of guidelines; in the best cases, leaving these considerations as future work or recommendations for successive research. Yet, despite their potential to inform the design of interactive systems, guidelines still can be confusing, contradictory, obsolete (due to the technology advances), or just too many and thus, difficult to handle \cite{abascal2004use}. 

Hence, this work objectives are to critically investigate existing touchscreen design guidelines and identify weaknesses and shortcomings in this field. We aim at making the guidelines more useful for designers and developers, supporting them in their understanding of the relevance of each guideline, its accessibility, clarity, and validity. We believe that increasing the adoption of guidelines would result in more usable and accessible touchscreen applications, and thus, benefit older adults and the general population.

\section{Methods}
The systematic review was done according to the PRISMA framework \cite{moher2009preferred}. Three researchers (independently) screened titles, abstract and full text of articles following the procedure below.

\textbf{Identification.} Search for articles that include keywords associated to either ageing, guidelines, and touchscreen in their title or abstract. The search was limited to articles published in relevant journals or conference proceedings in the areas related to HCI and ageing, published between 2005 and 2016, and written in the English language.\\
\textbf{Screening.} Screen identified articles to evaluate if the article contains design guidelines for touch devices or applicable to touch devices. Articles were tagged with \textit{Yes, No, and Maybe} to mark if they contained guidelines. Resolve disagreements with the other researchers with respect to articles tagged with \textit{Maybe} in face-to-face discussions.\\
\textbf{Eligibility.} Evaluate in details each article and extract the proposed guidelines if applicable. During this step, articles were also tagged with \textit{No} if the proposed guidelines were too general or confusing.\\
\textbf{Include.} Extract from each article the details of the studies that either conducted to the proposed guidelines and/or validated them. Each study was evaluated based on its formulation of outcome measures, study design, population and source.
\\The results of each step can be seen in Figure \ref{figure:collage}.

Design guidelines were extracted from each of the selected articles using a standardized data coding form, which resulted into a table of study records that were further reviewed by coders. Coding parameters were the following: date of publication, authors, short summary, type of ability decline and its screening methods if any, type of target touchscreen device, pre-studies (that guided the creation/definition of the proposed guidelines) and post-studies (that either applied the guidelines, or validated them) including data about subjects (size, age, percentage of females), format of user study (group or individual), mode of assessment (technical or non-technical), and presentation of the final design recommendations. 

\section{Results}
The review resulted in a set of 204 design guidelines to support the design of solutions targeting declines of abilities related to ageing. This list was analyzed, filtered, further integrated, and transcribed to create an operational version of guidelines. The final list of included papers and respective guidelines is available on the website at \url{http://happy.mateine.org/design4all}.

Despite the valuable recommendations in the resulting guidelines, the compilation process uncover some shortcomings in terms of formulation, poor organization and structuring, and reliability of findings. We discuss those issues below.


\subsection{Clarity of guidelines}
Extracting design guidelines from the list of included papers was a complicated task. The majority of papers presented the guidelines in the form of discussions, using unclear formulation (sometimes simply incomprehensible), and so making it difficult to assess whether certain findings could be indeed defined as design guidelines. 



These issues were addressed in our process with an extraction phase performed by three independent researchers, followed by face-to-face discussions to resolve disagreements (a considerable number). To further avoid ambiguity, we contacted the original authors of the articles asking for feedback on the identified guidelines, and some of them rephrased or corrected the design recommendations. Taken together, this work resulted into the final list of included guidelines.

The above findings indicate missed opportunities. Making guidelines and contributions difficult to identify and consume prevent the uptake of recommendations by the larger community. The use of standard reporting formats\footnote{CONSORT and PRISMA are examples of reporting standards for clinical trials and systematic reviews} for reporting and the development of knowledge bases could help address this issue and provide benefit to the whole community. 

\subsection{Guideline organization}
The heterogeneity of the elderly population makes it difficult to apply the same design guidelines on this entire population. Thus, guidelines should ideally define the specific target population as to clearly communicate its scope and applicability. 

The majority of the articles in our review (58\% of them) were found to define older adults using a chronological approach (by age), usually as people older than 65 years \cite{carmien2014elders} or in average from 60 to 80 years with only one work explicitly focusing on the 'oldest old' (80+) population group \cite{neves2015my}. 
The rest of the studies adopted a functional approach and focused on ageing related ability declines, which could be considered as a way of targeting a diversity of individual ability declines of older adults \cite{sarcar2016towards}. These papers focused on one or more abilities, for example, studying interaction with touch devices by older adults with aphasia \cite{martin2013computerized}. 

In addition, just half of all studies used validated screening instruments in order to identify the presence of ability declines in older adults, the most common of them being visual acuity measuring tests, for example, Snellen eye chart \cite{darroch2005effect}.

According to these findings, in our review we adopted a classification based both on the taxonomy of ageing related ability declines and design categories, following the above recommendation. Manual coding of guidelines has been proven quite complicated, requiring not only coding by three experts but feedback from the authors. We believe that the adoption of a standard classification system in the reporting of guidelines would greatly improve the access and increase adoption.

\subsection{Reliability} 
As for the quality of identified findings, we have considered validation procedures described in selected studies. They included experiments and user studies with older adults \cite{mi2014heuristic}, as well as analysis of findings with comparison to existing literature. However, validation of proposed guidelines has been stated in about 45\% of works only, which represents a rather disappointing trend. This finding raises awareness of the need of further experimental investigations in order to determine the trustworthiness and efficiency of existing guidelines and providing an operational framework for new reliable design recommendations generation.

These findings complement those mentioned earlier and emphasize not only the need of having easier and clearer access to the best design practices for developing touchscreen applications for older adults, but also the need of a more structured approach in their categorization and validation.

\section{Future work} 
Considering that software designers and developers do not always have a realistic picture of the abilities and preferences of older adults, usually treating them as a homogeneous group that is affected by a set of physical and cognitive declines \cite{ijsselsteijn2007digital}, above discussed findings suggest several courses of action.

Following this work, we are currently depicting our findings and developing a website (\url{http://happy.mateine.org/design4all}) as a repository of guidelines derived from our review. We believe, it would allow researchers and developers to apply and consult the guidelines while developing touchscreen application or conducting studies for and with older adults. Despite that, further studies could usefully explore if existing touchscreen design guidelines are valid in covering abilities of heterogeneous groups of ageing population. 

Another practical implication we try to tackle is categorization of guidelines by various degrees, types, and combinations of ability declines, as well as usability problems older adults may face. This could be a step to support designers and developers in creating products adapted to the ageing population and understanding its diversity, as well as providing a platform to support contributions into this area.
\balance{} 
\section{Acknowledgements}
This project has received funding from the EU Horizon 2020 research and innovation programme under the Marie Sk\l{}odowska-Curie grant agreement No 690962. This work was also supported by the ``Collegamenti'' project funded by the Province of Trento (l.p. n.6-December 13rd 1999)
\bibliographystyle{SIGCHI-Reference-Format}
\bibliography{sample}
\end{document}